\def\year{2020}\relax
\documentclass[letterpaper]{article} 
\usepackage{aaai20}  
\usepackage{times}  
\usepackage{helvet} 
\usepackage{courier}  
\usepackage[hyphens]{url}  
\usepackage{graphicx} 
\usepackage{comment}
\usepackage{graphicx}  
\usepackage{latexsym}
\usepackage{pgfplots}
\usepackage{algorithm}
\pgfplotsset{width=4.3cm,compat=1.7}
\usepackage{pgfplotstable}
\usepackage[caption=true]{subfig}
\usepackage{enumitem}
\usepackage{booktabs}
\usepackage{multirow}
\usepackage{float}
\usepackage{array, makecell}
\usepackage{boldline}
\usepackage{amsfonts}
\usepackage{amsmath}
\usepackage{textpos}

\title{On Predicting Personal Values of Social Media Users using Community-Specific Language Features and Personal Value Correlation\thanks{To be appear in the proceedings of ICWSM 2021}}
\author{Amila Silva\textsuperscript{\rm 1}, Pei-Chi Lo\textsuperscript{\rm 2}, and Ee-Peng Lim\textsuperscript{\rm 2}\\ 
\textsuperscript{\rm 1}School of Computing and Information Systems, The University of Melbourne,\\
Melbourne, Australia\\
\textsuperscript{\rm 2}
School of Information System, Singapore Management University,\\
Singapore\\
amila.silva@student.unimelb.edu.au, \{pclo.2017@phids., eplim@\}.smu.edu.sg
}
\begin{document}

\maketitle
\begin{abstract}
Personal values have significant influence on individuals' behaviors, preferences, and decision making.  It is therefore not a surprise that personal values of a person could influence his or her social media content and activities. Instead of getting users to complete personal value questionnaire, researchers have looked into a non-intrusive and highly scalable approach to predict personal values using user-generated social media data. Nevertheless, geographical differences in word usage and profile information are issues to be addressed when designing such prediction models. In this work, we focus on analyzing Singapore users' personal values, and developing effective models to predict their personal values using their Facebook data. These models leverage on word categories in Linguistic Inquiry and Word Count (LIWC) and correlations among personal values. The LIWC word categories are adapted to non-English word use in Singapore. We incorporate the correlations among personal values into our proposed Stack Model consisting of a task-specific layer of base models and a cross stitch layer model. Through experiments, we show that our proposed model predicts personal values with considerable improvement of accuracy over the previous works. Moreover, we use the stack model to predict the personal values of a large community of Twitter users using their public tweet content and empirically derive several interesting findings about their online behavior consistent with earlier findings in the social science and social media literature.
\end{abstract}

\noindent \noindent \section{Introduction}

\begin{figure}[t]
\centering
\includegraphics[width = 0.75 \columnwidth]{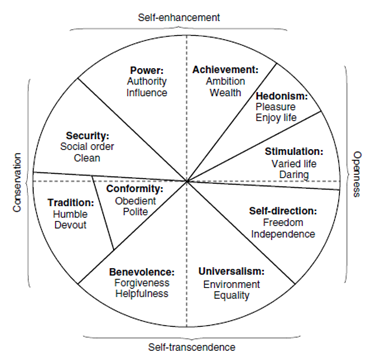}
\caption{Circular Structure of Schwartz's Personal Values ~\cite{maio2010mental}}
\label{fig:theory}
\vspace{-4mm}
\end{figure}

\textbf{Motivation}. \textit{Personal values} (or simply, values) express what is most important to people in life. Every individual holds personal values (e.g., achievement, security, benevolence) with varying degrees of importance. A particular personal value may be very important to one person but unimportant to another. Personal values influence a person's attitude and behavior. Although there are various approaches \cite{hofstede1984culture,braithwaite1985structure,inglehart1997modernization} to measure personal values, the one proposed by Schwartz in~\cite{schwartz1992universals} is widely used by psychology researchers. Schwartz's theory of basic human values \cite{schwartz2012overview} defines 10 specific personal values which can be measured by a specially designed questionnaire. These personal values are: Power, Achievement, Hedonism, Stimulation, Self-Direction, Universalism, Benevolence, Tradition, Conformity, and Security (see Fig.~\ref{fig:theory}). These personal values are further grouped into five higher order personal values, namely: \textit{Self-enhancement}, \textit{Hedonism}, \textit{Openness}, \textit{Self-transcendence}, and \textit{Conservation}.  The questionnaire includes 56 items \cite{schwartz2003proposal} to measure the value dimensions\footnote{To the best of our knowledge, other works also focus on analyzing and predicting high-order personal values \cite{mukta2017identifying,chen2014understanding}.}. However, it is costly and privacy-intrusive to get people to complete the questionnaire. Researchers thus seek other more scalable alternative approaches to obtain personal values. Among these approaches, prediction of personal values from users' social media content is very promising and has been studied in only a few works \cite{boyd2015values,chen2014understanding}. Because, such descriptors related personality could be subsequently served as features for the downstream applications like personalized recommendations~\cite{kern2019social}. 

There also exists some software products to predict personal values based on content\footnote{https://www.ibm.com/watson/services/personality-insights/}, they are based on models trained on labeled users from specific geographical region and culture. These models may not predict accurately for another community of users due to: (a) different personal value profile distribution among new users compared with those used in training the models; and (b) some word use patterns of new users very different from that of training data. While the above limitations are well understood, there has not been much research to illustrate the impact of community specific language and personal value profile distribution to personal values prediction using user-generated content.  

\textbf{Research Objectives.} This work aims to show that accurate value prediction has to leverage on word usage patterns specific to the region the users come from.  A value prediction model trained using the content from another region may yield poorer performance when applied to the target region even the two regions have large lexicon overlap.  At times, the prediction accuracy may not be high enough for data science studies on a large social media user population. The important research questions to address in this work are therefore: (a) how to cope with regional differences in word usage patterns? (b) how to develop accurate models to leverage features and task knowledge? (c) how to deploy value prediction models to gain insights about a large social media user population?    

To answer the above questions with concrete illustration, we focus on personal values of users from Singapore which has an ethnic composition of 76\% Chinese,  15\% Malays, 7\% Indians and others. Most Singapore users are asians who may have personal value profiles different from users from the US and European communities. The languages used in the Singapore community has major content differences which may render the previous models less ineffective. To study the personal value profiles of Singapore users and to develop personal value prediction models using their social media data, we have collected the Facebook content generated by a group of Singapore users.  According to Alexa at the time of this study, Facebook is the most visited social media site in Singapore.  The users in this group also completed the Schwartz's personal value questionnaire so as to have their personal values determined.  

In developing the prediction models, we explore several novel ideas.  The first idea is to use a Singapore variant of \textit{Linguistic Inquiry and Word Count} (LIWC) instead of the original LIWC to derive features adapted to the linguistic characteristics of Singapore users.  LIWC consists of word categories that characterize the linguistics profiles of users. The second idea leverages on the positive and negative correlations between personal values according to the \textit{circular structure of Schwartz's Personal Values} \cite{schwartz1992universals} as shown in Fig.~\ref{fig:theory}. This work proposes a method to exploit the circular structure of personal values to boost their prediction accuracy. Both idea have not been studied in any previous works on personal value prediction. 

In this work, we further illustrate a scalable way to analyse online behavior of a large group of Twitter users using our personal value prediction model trained using Facebook data. Despite the analysis involves different social media data, we empirically show that the predicted labels of our model uphold the inter-relationships among personal values \cite{schwartz1992universals}, highlight some hypotheses about user online behaviors in Twitter, and compare them with those reported in related works~\cite{marshall2018intellectual,jin2013peeling,huberman2008social}.

\textbf{Contributions.} In summary, the following are the novel contributions of this work:
\begin{itemize}
\item We introduce a novel dataset collected for Singapore Facebook users with the ground truth labels of Schwartz's personal values\footnote{\label{note1}The anonymous dataset is shared via \url{shorturl.at/fhjS9}.}  Singapore users represent a unique user community with its own linguistic characteristics. To our best knowledge, we are the first to analyze and predict personal values using a community-specific LIWC incorporating these linguistic characteristics. This is known as the \textit{Singapore-LIWC} or S-LIWC.
\item We propose a new personal value prediction model known as \textit{Stack Model} to improve the prediction accuracy of the personal values. The key idea is to exploit the circular structure of Schwartz's personal values. This Stack Model is a two-layer model which supports features derived from both LIWC and S-LIWC.  The model yields accuracy higher than most of the earlier state-of-the-art prediction models.
\item Instead of using social media post content only, we investigate user representation using features from Facebook profile information (i.e., interests and groups). Empirically, we show that the profile features are strong features to predict personal values. 
\item We conduct a data science study of the personal values of a large community of Singapore Twitter users (85,000 users) who share their tweet content publicly, and correlate the personal values with their online behavior. In this study, our proposed stack model discovers interesting online user behavior with specific personal values. This is a major breakthrough extending the study of personal values to a large user community.
\end{itemize}

\section{Related Works}
\label{sec:related}
We divide the related works into two categories, namely: (a) \textit{personal value prediction}; and (b) \textit{personality prediction}.

\textbf{Personal Value Analysis and Prediction.} Personal values have been studied in the context of decision making and personal interests. In~\cite{verplanken2002motivated}, the effect of individual's personal values over their decision process is analyzed. Hsieh et al. studied the relationship between personal values and personal interests \cite{hsieh2014you}. There are a few previous efforts \cite{chen2014understanding,boyd2015values,maheshwari2017societal} to analyze the relationship between personal values and word usage. In \cite{chen2014understanding}, Chen et al. uses Reddit as the social media platform to collect user generated online context while Boyd explicitly asks users to produce content \cite{boyd2015values}. LIWC word categories and modeled topics are used in \cite{chen2014understanding} and \cite{boyd2015values} respectively to capture content features. Both works show the correlation analysis between word usage and personal values. Chen et al. further analyzes the prediction potential of personal values using simple binary classification models such as Logistics Regression, SVM, and Naive Bayes. Nevertheless, there has been little work on personal value prediction for user communities using region/culture-specific languages. 

Personal value prediction using Facebook data is new. In addition to the word content in posts, Facebook profiles consist of other user generated profile data such as interests, group affiliations, and activities. These profile data however have not yet be used as features for personal values prediction in similar previous works \cite{chen2014understanding,boyd2015values}. There also has not been any other previous works on large-scale data science study of personal values and online behavior.

\begin{table*}
\centering
\scriptsize
\caption{Comparison between the statistics of our Facebook dataset and Reddit dataset \protect\cite{chen2014understanding} (ST=Self-Transcendence, SE=Self-Enhancement, CO=Conservative, OC=Openness to Change, HE=Hedonism. Significant correlations are shown in boldface.)}
\label{tab:stat}
\begin{tabular}{|lV{3}r|r|r|r|r|rV{3}r|r|r|r|r|rV{3}}
\hline
& \multicolumn{6}{cV{3}}{Our Dataset} &\multicolumn{6}{cV{3}}{Reddit Dataset \cite{chen2014understanding}}\\
\hline
 &  \multirow{ 2}{*}{Mean} & \multirow{2}{*}{Std Dev} & \multicolumn{4}{cV{3}}{Correlations}& 
 \multirow{ 2}{*}{Mean} & \multirow{2}{*}{Std Dev} & \multicolumn{4}{cV{3}}{Correlations}\\
\cline{4-7}\cline{10-13}
& & & SE & CO & OC & HE&&& SE & CO & OC & HE\\
\hline
ST & 0.30 & 0.50& \textbf{-0.63}&0.10&\textbf{-0.28}&\textbf{-0.54} 
& 0.85 & 0.63& \textbf{-0.58}&-0.20&-0.07&\textbf{-0.23}\\
\hline
SE  & -0.63 & 0.77& &\textbf{-0.32}&0.09&\textbf{0.33}
& -0.50 & 0.73& &\textbf{-0.25}&-0.19&-0.02\\
\hline
CO & -0.31 & 0.58& &&\textbf{-0.65}&\textbf{-0.40}
& -0.86 & 0.66& &&\textbf{-0.66}&\textbf{-0.34}\\
\hline
OC & 0.04 & 0.80& &&&\textbf{0.27}
& 0.44 & 0.60& &&&\textbf{0.61}\\
\hline
HE & 0.01 & 1.18& &&&
& 0.26 & 0.95& &&&\\
\hline
\end{tabular}
\vspace{-4mm}
\end{table*}


\textbf{Personality Prediction.} Prediction of other psychological attributes such as personality \cite{golbeck2011predicting}, and dark triad personality traits \cite{sumner2012predicting} based on word usage have been studied in recent years. There were several efforts \cite{grankvist2015personality,parks2015personality} to analyze the relationship between personal values and personality traits. In addition, researchers have studied specific human behavior, thinking pattern or preference using personal values or personality traits \cite{md2017predicting,hsieh2014you,verplanken2002motivated}.  Mukta et al. analyzed the prediction potential of individuals' movie genre preferences using personality and personal values \cite{md2017predicting}. Most of the previous efforts generally use content features with simple classification models to perform prediction \cite{chen2014understanding,sumner2012predicting}. To the best of our knowledge, none of them exploits the circular structure among personal values to improve the prediction potential of personal values. Golbeck et al. proposed a few discrete features extracted from Facebook profile information for predicting personality \cite{golbeck2011predicting}. 

\section{Datasets and Data Analysis}
\label{sec:dataset}

In this paper, we use two datasets covering users from the same community.  The first dataset is a small Facebook dataset with ground truth personal value labels for training prediction models.  The second dataset is a large Twitter dataset covering more than 80K users and their tweet and social network data. The Twitter dataset allows us to predict the personal values of the target Twitter user community and to conduct a large-scale study of user behavior for users of different personal values.

\textbf{Facebook Dataset Construction. }
Social media datasets with ground truth personal value labels for research are generally not available. We therefore construct our own dataset by recruiting 125 undergraduate students (42 males and 83 females) 
to contribute their Facebook data and personal values labels by completing the Schwartz Value Survey (SVS) \cite{schwartz1992universals}. Each participant received a small monetary reward for the participation. 

The Schwartz Value Survey (SVS) is currently the most widely used personal value assessment instrument \cite{caprara2006personality,chen2014understanding,boyd2015values,md2017predicting}. This survey requires participants to rate the importance of 56 value items as guiding principles in their life within the scale from -1 to 7. To remove individual differences in rating, we subtract every user's original ratings across all personal values by the average of all ratings of the user. In this way, all users after the above adjustment share the same average rating of 0. This is the same technique used in previous works~\cite{chen2014understanding} to preprocess the ratings, thus we follow the same for the comparison purposes. 

Along with the survey, participants shared with the research team their Facebook profile archives, which consist of Facebook posts and other profile information such as gender, preferences, and network details. The 125 users together have a total of 383,335 posts which were posted within the time span from the 4th quarter of 2007 to 1st quarter of 2018. They are required to remove sensitive posts before handing their data to the research team. The least and most prolific users have 12 posts and 20971 posts respectively (Avg. number of posts per user = 3067). All the users had been members of Facebook for at least 18 months at the end of year 2017. Personal values are generally considered as rather stable broad psychological attribute \cite{rokeach1973nature,schwartz1992universals} compared to other similar attributes such personality traits. Hence it is still valid to analyze individuals' personal values using their older posts. 

To check the \textit{internal consistency} of survey results, we calculated Cronbach's alpha \cite{cronbach1951coefficient} for each higher-order value dimension. 
The alpha values range from 0.73 to 0.86 (even higher) for the 5 higher-level personal values. These results are considered quite high  and far better than that derived from the dataset used in \cite{chen2014understanding}.  

To further ensure that the dataset is well collected, we compare the statistics of our dataset with that of a dataset used in \cite{chen2014understanding}. According to Table~\ref{tab:stat}, most of the statistical figures of our dataset are similar to that of dataset in \cite{chen2014understanding}. In both datasets, the mean value for Self-Enhancement and Conservative are negative and it is positive for the rest. Highest variance between users is observed for the dimension of Hedonism in both datasets. On the other hand, Table~\ref{tab:stat} also reveals that our dataset has low mean values for all the personal value dimensions except for Conservative compared to the previous dataset \cite{chen2014understanding}. This suggests that Singapore users are highly driven by the goals like acceptance of tradition and customs, safety and stability of society, and relationship. This also highlights the importance of this research, which is done for a specific user community to yield more accurate prediction results and findings.

Most of the correlations between value dimensions are also similar between the two datasets in terms of sign and magnitude (e.g., Self-Transcendence vs Self-Enhancement, Conservative vs Openness to Change, and Conservative vs Hedonism). The correlation values, which are different between our Facebook dataset and the \cite{chen2014understanding}'s dataset can also be explained using the circular structure of the Schwartz's personal values. For example, our dataset shows a significant positive correlation between self-enhancement and hedonism while it is a weak negative correlation in \cite{chen2014understanding}'s dataset. Between the two, positive correlation is more consistent with the personal values theory as Self-Enhancement and Hedonism are closer to each other in the circumplex structure. Hence, we can conclude that our dataset is reasonably well-collected and it complies with the circular structure of Schwartz's personal values, which will be exploited in our subsequently proposed model for accurate prediction. 

\textbf{Twitter Dataset Construction. }
To conduct a larger-scale study of personal values for users from the Singapore community, we collected \textit{public tweets} posted by 85,308 Twitter users based in Singapore during the 6 months period from January 2017 to June 2017. The dataset was constructed by first identifying a seed set of well known Twitter users based in Singapore. By crawling their following links, we reached out to other users who are also based in Singapore. We repeated the steps until the user set does not increase by size. In addition to the tweets, the Twitter network between these users (i.e., followers, friends) was also extracted. All these post and social network data are subsequently used in our user behavioral study. The descriptive statistics of this dataset are shown in Table~\ref{tab:twitter_data}.

\begin{table}
\scriptsize
\centering
\caption{Statistics of the Twitter dataset}\label{tab:twitter_data} 
\begin{tabular}{|c|l|c|c}
\hline
\multirow{2}{*}{\makecell{Post \\Statistics}}&Total Tweets                         & \multicolumn{2}{|c|}{9,499K} \\ \cline{2-4}
&Average Tweets per user              & \multicolumn{2}{|c|}{111}  \\ \hline \hline
\multirow{2}{*}{\makecell{Network \\Statistics}}& Network Type & Friend & \multicolumn{1}{c|}{Follower}\\ 
\cline{2-4}
& Number of edges                      & 1,585,060                                                  & \multicolumn{1}{c|}{2,988,157} \\ \hline
\end{tabular}
\vspace{-4mm}
\end{table}


\section{Proposed Prediction Models}
\label{sec:model}

\subsection{Task Definitions} 

By completing the SVS questionnaire, every user $u$ has a ground truth score for each personal value $p$ denoted by $v_{u,p}$.  In this paper, we focus on high-order personal values, namely: Self-Transcendence, Self-Enhancement, Conservation, Openness-to-Change, and Hedonism.  Recall that $\{ v_{u,p}\}$'s have been normalized to remove individual's leniency (or stringency) in ratings.

We formulate personal values prediction as a classification task similar to the previous work \cite{chen2014understanding}. 
We divide users into two equal-sized groups for each personal value $p$: \textit{top $K\%$ users} and \textit{bottom $K\%$ users} by $v_{u,p}$. The goal of the prediction task is to determine the group label of any new user as accurate as possible. 

In the previous works \cite{chen2014understanding,sumner2012predicting}, $K=50$ was used, i.e., a mid-split classification. 
For each value dimension, the top 50\% users with highest ground truth personal values are labeled as \textbf{positive}, and the rest as \textbf{negative}. To offer a buffer in between the positive and negative users in the prediction task, we also try $K=40$. In this case, the top 40\% and the bottom 40\% users are labeled as positive and negative instances respectively, and the mid 20\% users are not used. 

\subsection{Feature Selection}

\textbf{Post features using LIWC word categories.} To solve the classification task, we need to derive relevant features from users' social media data.  Following the earlier work \cite{chen2014understanding}, we utilize all 90 word categories in the Linguistic Inquiry and Word Count (LIWC) \cite{pennebaker2015development} as features. These word categories are used to investigate individuals' beliefs, thinking patterns, social relationships, and personalities. For each LIWC word category, we obtain a feature score by the total frequencies of words from the word category found in the user's own content postings (also known as \textbf{post features}). 

\textbf{Post features using S-LIWC word categories.} For the Singapore's user community, a English-based creole language known as \textit{Singlish} is used widely to generate the social media content.  Unlike standard English, Singlish incorporates words and lexical rules from Chinese, Chinese dialects, Malay, and even Indian languages.  We therefore extend LIWC to incorporate Singish words so as to leverage on Singlish word features. For example, the Singlish sentence ``the question is very chim'' carries the same meaning as ``the question is difficult''. The word ``chim'' originates from a Chinese dialect.  In other words, one can find both English and non-English words co-exist in Singlish.  One can therefore exploit the similar context of similar words in the Singlish corpus to create a Singlish variant of LIWC known as \textbf{S-LIWC} (XYZ)\footnote{Reference is not provided due to double-blind.} 

In S-LIWC, the key idea is to use a Word2vec word embedding model \cite{mikolov2013distributed} trained on a corpus comprises of around 150,000 Singapore tweets.  With the learned model, Singlish words sharing similar context with words found in LIWC word categories are determined.  For example, if ``chim'' and ``difficult'' (which is a seed word in the LIWC negative emotion word category) are found to be close to each other in the embedding space.  One could therefore include ``chim'' as the Singlish word in the corresponding word category. 

The top $q$ closest words ($q$ was set to 10 empirically) for each LIWC seed word in the embedding space are then selected and added to the initial candidate word list of the seed word. As antonyms may share also similar context, they are removed using a logistic regression classifier trained to classify synonym-antonym relationship. This classifier is trained on the known synonyms and antonyms sets of LIWC seed words from Oxford Dictionary API\footnote{https://developer.oxforddictionaries.com/}. 
After removing all non-synonyms from the candidate word list using the classifier, a total of 9,640 distinct words are added to the various LIWC categories which form the S-LIWC. More details about the construction and evaluation S-LIWC can be found in (XYZ). 

{\bf Profile features.} Other than Facebook content posts, we also explore other non-textual behavioral user data as features. We found that each Facebook user profile offers information about user's  interests, activities, and groups which come with textual content. 
Similar to content posts, we extracted words from the these profile text and derive LIWC (and S-LIWC) word category scores as profile features.

\begin{table}
\scriptsize
\vspace{-2mm}
\centering
\caption{Coverage of LIWC and S-LIWC words in our Facebook Dataset}
\label{tab:coverage}
\begin{tabular}{|l|c|c|}
\hline
    & Posts  & Profile \\ 
\hline \hline
\# of unique words & 107,957 & 6021 \\
\hline
\# of unique LIWC words & 37,174 (34.43\%) & 2773 (46.06\%)  \\
\hline
\# of unique LIWC + S-LIWC words & 38,846 (35.98\%) & 3038 (50.46\%) \\ 
\hline
\end{tabular}
\end{table}

\textbf{Word coverage.} Table~\ref{tab:coverage} shows the coverage of LIWC and S-LIWC words in our Facebook dataset.  It shows that an additional 1672 and 265 unique S-LIWC words have been found in content posts and profile respectively. These include Singlish words such as \textit{``lah'', ``la'', ``xuan'',} and \textit{``wan''}, which are useful to analyze individuals' personal values. This observation further signifies the importance of having a community specific LIWC dictionary.

\begin{figure}
    \centering
    \includegraphics[width = 1.05\linewidth]{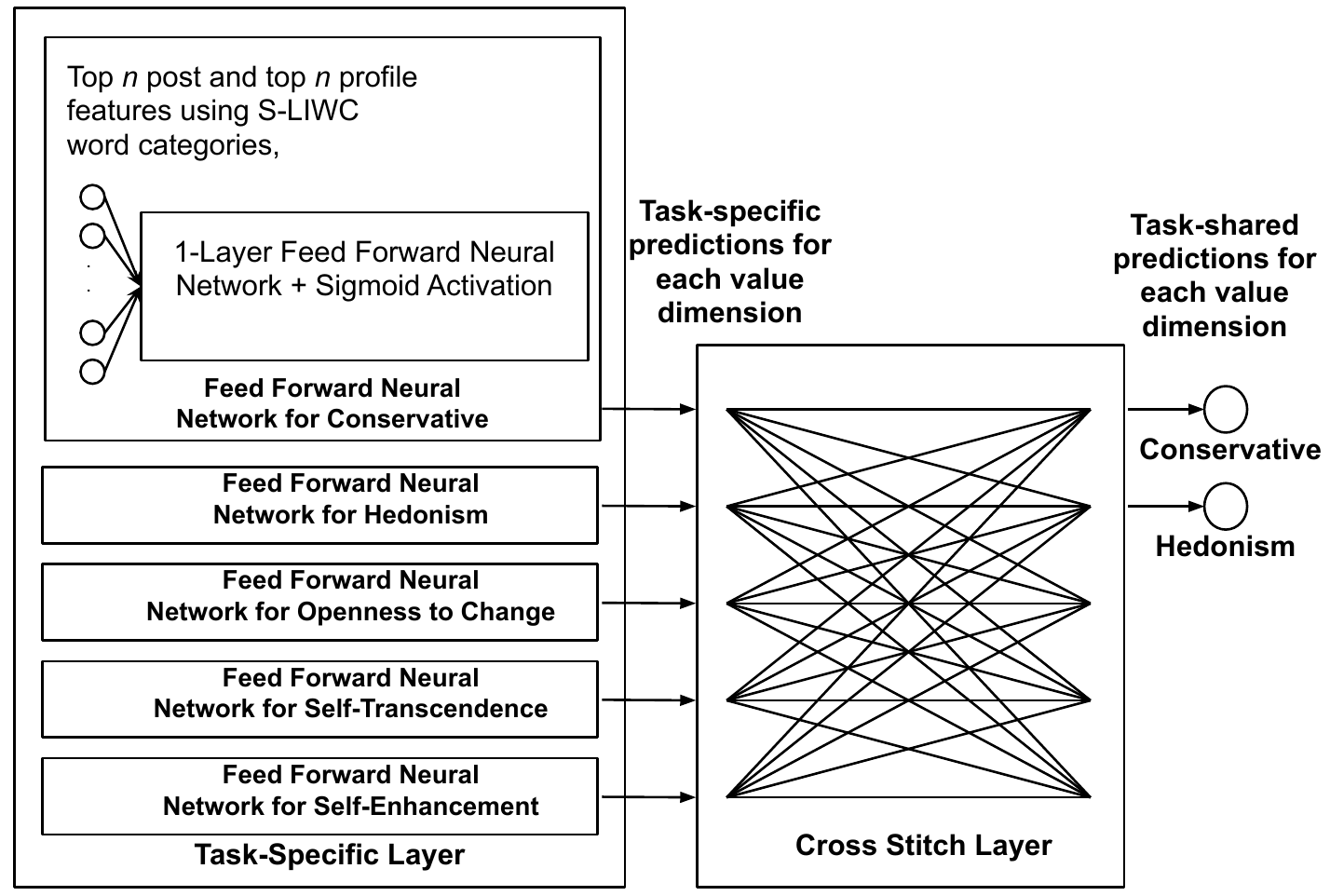}
  	\caption{Proposed Stack Model}
    \label{fig:stack}
    \vspace{-4mm}
\end{figure}

\subsection{Proposed Models}

Our proposed prediction models consists of a few base models which serve as the baselines.  We further propose a \textit{stack model} placing a cross stitch unit over a set of neural models one for each personal value.  In the following, we describe both the base and stack models.

\textbf{Base Models.} With the selected features and ground truth user labels, we train base models for  different personal values using Logistic Regression (LR).  We leave out the popular Support Vector Machine (SVM) as our experiment results show its performance is not better than LR. We also leave out more advanced neural models as our experiments (not reported in this paper) found out that many of these models could not perform well due to the small dataset. For each personal value, we train 6 base models using the following feature settings: (a) post features using LIWC, (b) profile features using LIWC, (c) both post and profile features using LIWC, (d) post features using S-LIWC, (e) profile features using S-LIWC, and (f) both post and profile features using S-LIWC. In each setting, we only use top $n$ post and top $n$ profile features, selected based on a feature selection strategy mentioned in Section~\ref{sec:evaluation}, as we empirically observe that using all the available features could lead to overfitting the model given our small dataset.

\textbf{Stack Model.} 
Unlike the previous efforts \cite{chen2014understanding,sumner2012predicting} using separately prediction models for different personal value dimensions, we propose the Stack Model which exploits the significant correlation in between value dimensions as shown in Table~\ref{tab:stat}.
In this model, the prediction models for value dimensions are learnt together using a multi-task learning approach. As shown in Figure~\ref{fig:stack}, the stack model has (a) a task specific layer which consists of a 1-layer feed forward neural network with sigmoid activation for each value dimension; and (b) a cross-stitch layer to combine output values from the task-specific layer and supervise how much sharing is needed among related tasks~\cite{misra2016cross}.  We further elaborate this model below.

\textbf{(a) Task-specific layer.} Since our dataset is rather small, our task-specific layer consists of one feed forward neural network for each personal value dimension to avoid overfitting.  Each feed forward neural network takes S-LIWC word categories as input features and returns prediction of a specific value dimension. Instead of feeding all S-LIWC features as input which may lead to overfitting even with strong regularization, we only use top $n$ post features and/or top $n$ profile features (using S-LIWC word categories) identified using a feature selection strategy (see Section~\ref{sec:evaluation}) for each neural network. We apply sigmoid activation function to keep the predictions in the range of [0,1]. Another softmax layer is not required here to normalize the prediction outputs across dimensions as one user may be assigned multiple value dimensions. Formally, let $X_{u,p} \in \mathbb{R}^{2n \times 1}$ be the selected top $n$ post and top $n$ profile features for personal value $p$ of user $u$. The task-specific prediction $\tilde{y}_{u,p}$ for personal value $p$ of user $u$ is predicted as:

\begin{equation}\label{eq:task-specific}
    \tilde{y}_{u,p}= \frac{\exp(A*X_{u,p} + b)}{\exp(A*X_{u,p} + b) + 1}
\end{equation}

where $A \in \mathbb{R}^{1 \times 2n}$ and $b \in \mathbb{R}^{1 \times 1}$ are the trainable linear projection weights and bias offset of the task-specific layer respectively.

\textbf{(b) Cross-stitch layer.} This layer takes task-specific prediction value for each value dimension and returns task-shared prediction value for each value dimension, which is calculated as the linear combination of task-specific predictions for different value dimensions. Formally, let $\tilde{Y}_{u}, \hat{Y}_{u} \in \mathbb{R}^{5 \times 1}$ denote the task-specific and task-shared value prediction for user $u$ respectively. The cross-stitch unit comprises a trainable linear matrix $Z \in \mathbb{R}^{5 \times 5}$ to capture the sharing between different tasks as follows.

\begin{equation}\label{eq:task-shared}
     \hat{Y}_{u} = \frac{\exp (Z * \tilde{Y}_{u})}{\exp (Z * \tilde{Y}_{u}) + 1}
\end{equation}

To learn the trainable parameters of the proposed stack model, we minimize following objective function using SGD, considering the task-specific and task-shared predictions into account and generalizing the loss function which considers only final predictions (task-shared predictions).

\begin{equation}\label{eq:loss-function}
\begin{split}
    L_{u,shared} &= Y_{u}\odot \hat{Y}_{u} + (1 - Y_{u})\odot (1 - \hat{Y}_{u})\\
    L_{u,specific} &= Y_{u}\odot \tilde{Y}_{u} + (1 - Y_{u})\odot (1 - \tilde{Y}_{u})\\
    L &= \sum_{u} L_{u, specific} + (1-\beta)*L_{u, shared}
\end{split}
\end{equation}

where $Y_{u} \in \mathbb{R}^{5 \times 1}$ denotes the actual ground truth labels for top $K$ prediction task of user $u$, and $\beta$ controls the weight given to task-specific and task-shared loss functions. $Y_u[v]=1$ if the user $u$ possesses personal value $v$, and $Y_u[v]=0$ otherwise.  $\odot$ represents the dot product operation.  We empirically observe that giving more weight to $L_{u, specific}$ during the initial training epochs leads to better generalization of the model. Hence, $\beta$ is set as $\exp (-m*training\;epoch)$, where $m$ (empirically set to $10^{-3}$) is a hyperparameter to control the slope of the $\beta$ value. $Z$ matrix is initialized as an identity matrix and all other parameters are initialized using a normal distribution. 

Empirically, we can observe that the weights learned in the final layer reflect the actual inter-correlations of the value dimensions, which is elaborated in Section~\ref{sec:evaluation}. Suppose our base model of conservative dimension predicts a low score for a given user who has open to change ground truth label. This predicted score, upon reaching the task-sharing layer of our stack model, contributes to an increase the score for openness to change (due to the opposite correlation between the two personal value dimensions) even when the base model of openness to change does not predict a high score for the same user. Hence, the stack model may still return a high score for openness to change. Likewise, our stack model exploits the inter-correlations between personal values to modify or reinforce the prediction results.

\section{Evaluation of Proposed Models}\label{sec:evaluation}

\begin{table*}[t]
\centering
\scriptsize
\caption{AUC ROC for top $K$\% Prediction using 5-fold cross validation (The best results are shown in boldface.)}\label{tab:results}
\scriptsize
\vspace{-2mm}
\begin{tabular}{|l|l||c||c||c|c|c|c|c|c|c|c|}
\hline
                        &    & LR (all) & IBM Watson & \multicolumn{3}{c|}{Base - LIWC}                                            & \multicolumn{3}{c|}{Base - S-LIWC}                                           & Stack LIWC & Stack S-LIWC\\ \hline
                        &    & Post                                                    & Post                                                  & Post  & Profile & Post $|$ Profile & Post  & Profile & Post $|$ Profile &   &     \\ \hline \hline
\multirow{5}{*}{$K = 50$} & CO & 0.487                                                    & 0.554                                                 & 0.711 & 0.632   & 0.711                                                     & 0.754  & 0.64    & \bf 0.795                                                     & 0.742&0.775  \\ \cline{2-12} 
                        & HE & 0.444                                                    & 0.47                                                  & 0.668 & 0.65    & 0.684                                                     & 0.698  & 0.696   & 0.633                                                     & 0.718&\bf 0.721  \\ \cline{2-12} 
                        & OC & 0.624                                                    & 0.525                                                 & 0.719 & 0.629   & 0.726                                                     & 0.747 & 0.608   & 0.758                                                     & 0.782&\bf 0.809  \\ \cline{2-12} 
                        & ST & 0.610                                                     & 0.589                                                 & 0.72  & 0.755   & 0.845                                                     & 0.714 & 0.775   & 0.849                                                     & 0.858&\bf 0.869  \\ \cline{2-12} 
                        & SE & 0.481                                                    & 0.480                                                  & 0.705 & 0.639   & 0.702                                                     & 0.754  & 0.744   & \bf 0.783                                                     & 0.726&0.777  \\ \hline \hline
\multirow{5}{*}{$K = 40$} & CO & 0.491                                                       & 0.573                                                 & 0.712 & 0.636   & 0.692                                                     & 0.708 & 0.568   & 0.704                                                     & 0.749&\bf 0.807  \\ \cline{2-12} 
                        & HE & 0.449                                                       & 0.569                                                 & 0.62  & 0.662   & 0.652                                                     & 0.688 & 0.668   & 0.664                                                     & 0.682&\bf 0.707  \\ \cline{2-12} 
                        & OC & 0.647                                                       & 0.541                                                 & 0.744 & 0.632   & 0.756                                                     & 0.76  & 0.668   & 0.756                                                     & 0.839& \bf 0.873  \\ \cline{2-12} 
                        & ST & 0.627                                                       & 0.608                                                 & 0.732 & 0.748   & 0.78                                                      & 0.708 & 0.726   & 0.82                                                      & 0.817& \bf 0.823  \\ \cline{2-12} 
                        & SE & 0.483                                                       & 0.454                                                 & 0.736 & 0.644   & 0.756                                                     & 0.768 & \bf 0.834   & 0.804                                                     & 0.802& 0.829  \\ \hline
\end{tabular}
\vspace{-4mm}
\end{table*}

\textbf{Experiment Setup.} We evaluate our proposed prediction models against two state-of-the-art baselines using our Facebook dataset, the only dataset we can use in this research.  The two baselines are:
\begin{itemize}
    \item IBM Watson Personality Insight API\footnote{https://www.ibm.com/watson/services/personality-insights/}: A well known personal value prediction package which has been used in many commercial settings. We report the predicted Schwartz's personal values returned by the API when it receives the post content from us.  
    \item LR (all): This follows basically the Chen's paper~\cite{chen2014understanding} and trains a logistic regression (LR) classifier (empirically observed that LR outperforms other classifiers such as naive Bayes, support vector machines, and decision tree) for each value dimension. All LIWC features are used as features in this model. The model however does not consider regional language differences and correlation between personal values. 
\end{itemize}

We implemented our proposed base and stack models using Scikit-learn  
and tensorflow\footnote{https://www.tensorflow.org/} respectively. We measure the prediction accuracy by Area Under ROC Curve (AUC ROC) as it allows the results to be comparable to that in \cite{chen2014understanding}. For the base models for the different personal value dimensions, we further select top $n$ LIWC word categories as features to represent each user. We compare the accuracy of the top $K$\% prediction task for $n \in \{5, 10, 15, 20, 30, 40, 85\}$. To select the top $n$ LIWC word category features, we considered different feature selection strategies like univariate correlation methods and, recursive feature elimination (RFE)\footnote{RFE recursively prunes away the least important features obtained using regression coefficients until the desired number of features is reached.} and found RFE performs better in most of the cases. We empirically found out that $n = 15$ yields the best performance for all the value dimensions. Hence, RFE and $n=15$ are used in all the subsequent experiments.  For the stack model, we have used base models using S-LIWC word categories as these features show better results than base models using LIWC word categories.

\textbf{Summary of Results.} 
Note that our baseline methods can only yield results based on post features only. As shown in Table~\ref{tab:results}, our proposed models outperform the baselines across different value dimensions for both $K=50$ and $K=40$. Our base and stack model results are far better than the random baseline, which returns exactly 0.5 for AUC.

Surprisingly, IBM Watson Personality API does not even outperform the random guess for Hedonism (0.470) and Self-Enhancement (0.480) value dimensions. Another interesting observation is that Chen's model (i.e., LR (all)) also perform poorly for this dataset, possibly due to its inability to generalize with limited data. This observation shows the effectiveness of our feature selection strategy.

\textbf{S-LIWC vs LIWC Features.}  Feature wise, our base models using S-LIWC outperform those using LIWC in most of the value dimensions.  The S-LIWC profile features perform particularly well for predicting Self-Transcendence, while the S-LIWC post features perform well for the rest. The concatenation of both post and profile feature sets further improves the prediction results considerably for most of the value dimensions. 

Our stack model using a two-layer neural architecture (involving base models using S-LIWC based features) gives the overall best performance.  This result is encouraging as it shows that both S-LIWC and personal values correlation contribute to the accuracy of personal values prediction.

Interestingly, profile features alone could often yield prediction accuracy comparable to post features when using S-LIWC (e.g., Base model using S-LIWC for predicting SE). This is not observed when using LIWC.  This is due to most of user profiles involving community-specific features.

\textbf{Stack vs Base Models.}  As shown in Table~\ref{tab:results}, our proposed stack model outperforms even the best-performing base models for most of the personal values. The improvements are especially significant in the Openness to Change (6.7\% better than Base-S-LIWC(Post+Profile)) and Hedonism (3.3\% than Base-S-LIWC(Post only)) value dimensions. To further analyze the results of the stack model, we report the feature weights derived from the final layer of our stack model in Table~\ref{tab:feaWeights}. It shows the inter-correlations between value dimensions learned by the stack model consistent with what we know from Schwartz's theory \cite{schwartz1992universals}.  The circular structure also implies that opposing values in the structure should have significant negative correlation. For example, Table~\ref{tab:feaWeights} shows significant negative weight assigned to the base model predicted conservative value $\hat{y}_{u,`conserv.'}$ which allows predicted conservative value in the first layer to contribute negatively to the final prediction of openness to change value. Only 73.9\% ($=\frac{4.28}{4.28+1.42+0.09}$) of the predicted score of base model for openness to change contributes to the final prediction.  24.5\% ($=\frac{1.42}{5.79}$) of the predicted openness to change score is determined by the predicted conservative value by the base model. Note that conservative value is opposite to openness to change in the circular structure of personal values. 

\begin{table}
\centering
\scriptsize
\caption{Feature weights derived from the stack models' cross stitch unit for the top 50\% prediction task}\label{tab:feaWeights}
\begin{tabular}{|l|c|c|c|c|c|}
\cline{2-6}
\multicolumn{1}{c}{} & \multicolumn{5}{|c|}{\bf Feature Weight} \\ 
\cline{2-6}
\multicolumn{1}{c}{} & \multicolumn{1}{|c|}{\bf CO} & 
\multicolumn{1}{c|}{\bf ST} & \multicolumn{1}{c|}{\bf OC} & \multicolumn{1}{c|}{\bf HE} & \multicolumn{1}{c|}{\bf SE} \\ \hline
{Conservative(CO)}                        & \textbf{4.98}                             & -0.03                                 & \textbf{-0.57}                         & 0                             & \textbf{-0.62}                                \\ \hline
{Self-Transcend.(ST)}                     & -0.18                            & \textbf{7.84}                         & 0                                   &  0               & \textbf{-0.57}                                \\ \hline
{Open. to Change(OC)}                    & \textbf{-1.42}                   & 0                                 & \textbf{4.28}                          & 0                             & -0.09                                 \\ \hline
{Hedonism(HE)}                              & \textbf{-0.71}                   & \textbf{-1.74}                                 & -0.24                                  & \textbf{4.41}                & 0                                     \\ \hline
{Self-Enhance.(SE)}                      & -0.32                              & \textbf{-1.36}                        & -0.48                                   & 0                         & \textbf{5.58}                        \\ \hline
\end{tabular}
\vspace{-4mm}
\end{table}



\begin{table*}[t]
\scriptsize
\centering
\caption{Top 15 Post and Profile Features based on LIWC and S-LIWC (The positively correlated features are \textbf{boldfaced}.)}\label{tab:topfea}
\resizebox{\textwidth}{!}{%
\begin{tabular}{|c|c||c|c||c|c||c|c||c|c|}
\hline
\multicolumn{2}{|c||}{\textbf{CO = Conservative}}              & \multicolumn{2}{c||}{\textbf{HE = Hedonism}}                  & \multicolumn{2}{c||}{\textbf{OC = Openness to Change}}        & \multicolumn{2}{c||}{\textbf{ST = Self-Transcendence}}             & \multicolumn{2}{c|}{\textbf{SE = Self-Enhancement}}      \\ \hline
LIWC                         & S-LIWC                     & LIWC                       & S-LIWC                      & LIWC                    & S-LIWC                         & LIWC                         & S-LIWC                         & LIWC                     & S-LIWC                    \\ \hline \hline
\multicolumn{10}{|c|}{\textbf{POST FEATURES}} \\ 
\hline
\textbf{1st pers plural}     & \textbf{1st pers plural}  & 1st pers singular          & \textbf{Adjectives}        & 1st pers singular       & 1st pers singular             & \textbf{1st pers plural}     & \textbf{1st pers plural}      & \textbf{Compare}         & \textbf{Interrogatives}  \\ \hline
Anger                        & Anger                     & 1st pers plural            & Compare                    & Adjectives              & \textbf{Negation}             & 3rd pers plural              & \textbf{2nd person}           & \textbf{Interrogatives}  & Female                   \\ \hline
\textbf{Sadness}             & \textbf{Sadness}          & 3rd pers plural            & \textbf{Anxiety}           & \textbf{Interrogatives} & Adjectives                    & Auxiliary verbs              & \textbf{3rd pers singular}    & Quantifiers              & \textbf{Discrepancies}   \\ \hline
Female                       & Female                    & \textbf{Negative emotion}  & \textbf{Friends}           & \textbf{Anxiety}        & \textbf{Interrogatives}       & Common adverbs               & 3rd pers plural               & Female                   & Biological Processes     \\ \hline
\textbf{Male}                & \textbf{Male}             & \textbf{Friends}           & Female                     & \textbf{Anger}          & \textbf{Anxiety}              & \textbf{Conjunctions}        & \textbf{Quantifiers}          & Biological Processes     & Body                     \\ \hline
\textbf{Body}                & \textbf{Body}             & Female                     & Male                       & Sadness                 & \textbf{Anger}                & \textbf{Regular verbs}       & Anger                         & Health                   & Health                   \\ \hline
\textbf{Health}              & \textbf{Health}           & Male                       & \textbf{Discrepancies}     & Insight                 & Sadness                       & Interrogatives               & \textbf{Sadness}              & \textbf{Ingesting}       & \textbf{Ingesting}       \\ \hline
Sexuality                    & Sexuality                 & \textbf{Discrepancies}     & \textbf{Tentativeness}     & \textbf{Discrepancies}  & Insight                       & Friends                      & Friends                       & \textbf{Reward focus}    & \textbf{Risk}            \\ \hline
\textbf{Achievement}         & \textbf{Achievement}      & \textbf{See}               & Health                     & \textbf{Certain}        & \textbf{Discrepancies}        & \textbf{Female}              & \textbf{Female}               & \textbf{Risk}            & Future focus             \\ \hline
Reward focus                 & Reward focus              & Health                     & \textbf{Sexuality}         & Power                   & Hear                          & Cause                        & Cause                         & Future focus             & \textbf{Informal speech} \\ \hline
\textbf{Risk}                & \textbf{Risk}             & \textbf{Sexuality}         & Risk                       & \textbf{Reward focus}   & \textbf{Sexuality}            & Discrepancies                & Discrepancies                 & \textbf{Informal speech} & \textbf{Swear words}     \\ \hline
Past focus                   & Past focus                & Risk                       & Home                       & Risk                    & Risk                          & See                          & See                           & Netspeak                 & Netspeak                 \\ \hline
\textbf{Home}                & \textbf{Home}             & \textbf{Past focus}        & Money                      & \textbf{Motion}         & \textbf{Motion}               & Work                         & \textbf{Feel}                 & Colons                   & Colons                   \\ \hline
Death                        & Death                     & Money                      & \textbf{Death}             & \textbf{Swear words}    & \textbf{Swear words}          & \textbf{Colons}              & \textbf{Biological Processes} & \textbf{Dash}            & \textbf{Dash}            \\ \hline
Swear words                  & Swear words               & Question marks             & \textbf{Quotation marks}   & \textbf{Parentheses}    & \textbf{Parentheses}          & \textbf{Quotation marks}     & \textbf{Quotation marks}      & Quotation marks          & Quotation marks          \\ \hline \hline

\multicolumn{10}{|c|}{\textbf{PROFILE FEATURES}} \\ 
\hline
\textbf{Conjunctions}        & 1st pers plural           & \textbf{1st pers singular} & \textbf{1st pers singular} & 1st pers plural         & \textbf{3rd pers singular}    & 3rd pers plural              & 3rd pers singular             & \textbf{1st pers plural} & \textbf{Auxiliary verbs} \\ \hline
Interrogatives               & \textbf{2nd person}       & \textbf{1st pers plural}   & \textbf{1st pers plural}   & \textbf{Adjectives}     & Impersonal pronouns           & \textbf{Impersonal pronouns} & \textbf{Impersonal pronouns}  & Compare                  & \textbf{Interrogatives}  \\ \hline
\textbf{Negative emotion}    & Auxiliary verbs           & 2nd person                 & \textbf{2nd person}        & Negative emotion        & \textbf{Conjunctions}         & Auxiliary verbs              & \textbf{Conjunctions}         & \textbf{Interrogatives}  & \textbf{Quantifiers}     \\ \hline
Anxiety                      & Conjunctions              & Impersonal pronouns        & 3rd pers singular          & Female                  & \textbf{Negation}             & \textbf{Negation}            & Negation                      & \textbf{Anxiety}         & Anxiety                  \\ \hline
\textbf{Female}              & \textbf{Anxiety}          & \textbf{Common adverbs}    & \textbf{Auxiliary verbs}   & \textbf{Discrepancies}  & Positive emotion              & \textbf{Sadness}             & Quantifiers                   & Cause                    & Sadness                  \\ \hline
Discrepancies                & \textbf{Family}                    & Negation                   & Conjunctions               & Tentativeness           & Anxiety                       & \textbf{Female}              & \textbf{Sadness}              & \textbf{Discrepancies}   & \textbf{Family}          \\ \hline
Tentativeness                & \textbf{Female}           & Positive emotion           & Interrogatives             & Certain                 & Anger                         & \textbf{Differentiation}     & Family                        & Differentiation          & \textbf{Cause}           \\ \hline
\textbf{Ingesting}           & Insight                   & Sadness                    & Sadness                    & percept                 & \textbf{Biological Processes} & Feel                         & \textbf{Male}                 & \textbf{Feel}            & \textbf{Discrepancies}   \\ \hline
Power                        & See                       & Cause                      & Female                     & Feel                    & Body                          & \textbf{Health}              & \textbf{Certain}              & Drives                   & Certain                  \\ \hline
\textbf{Reward focus}        & \textbf{Feel}             & Certain                    & \textbf{Health}            & Reward focus            & Health                        & \textbf{focuspresent}        & \textbf{See}                  & \textbf{Affiliation}     & Body                     \\ \hline
Death                        & Sexuality                 & \textbf{Feel}              & Risk                       & \textbf{Future focus}   & \textbf{Home}                 & Informal speech              & Health                        & \textbf{Power}           & Motion                   \\ \hline
Informal speech              & \textbf{Home}             & \textbf{Sexuality}         & \textbf{Future focus}      & Home                    & \textbf{Religion}             & \textbf{Assent}              & Achievement                   & \textbf{Nonfluencies}    & Informal speech          \\ \hline
Swear words                  & \textbf{Religion}                  & \textbf{Risk}              & Religion                   & Religion                & \textbf{Assent}               & \textbf{Semicolons}          & Nonfluencies                  & Colons                   & Swear words              \\ \hline
\textbf{Netspeak}            & Nonfluencies              & \textbf{Swear words}       & Assent                     & \textbf{Swear words}    & \textbf{Nonfluencies}         & Quotation marks              & \textbf{Semicolons}           & \textbf{Semicolons}      & \textbf{Netspeak}        \\ \hline
\textbf{Nonfluencies}        & \textbf{filler}           & Semicolons                 & \textbf{Nonfluencies}      & Semicolons              & Semicolons                    & Parentheses                  & Question marks                & \textbf{Question marks}  & \textbf{Nonfluencies}    \\ \hline
\end{tabular}}%
\end{table*}

\textbf{Top Feature Analysis. }Table~\ref{tab:topfea} lists the top 15 features of the base models using LIWC and S-LIWC for each value dimension. A positive (or negative) correlation between a word category and a value dimensions means that users high in the value dimension use words from the word category frequently (or rarely). We thus attempt to interpret some additional relationships captured by word categories with respect to the underlying goals of the value dimensions. 

For Self-transcendence, the top positively correlated S-LIWC word category features such as ``feel'', ``sadness'', and ``1st person plural'' are highly reasonable. Users high in self-transcendence adopt a people-oriented sensitive persona, which is inlined with the defining goals of benevolence and universalism values. The top negatively correlated S-LIWC features are word categories such as ``achievment'' and ``anger'', which are top positively correlated features for Self-Enhancement. This result further demonstrates the correlation between personal values, i.e., a negative correlation between Self-transcendence and Self-Enhancement. In contrast, top LIWC word category features includes a few function word categories such as adverbs and punctuation marks. 

For Conservative, we observe that top 15 features extracted from posts using S-LIWC and LIWC are similar. Only the profile features shows some differences between the top features from S-LIWC and LIWC. This shows that profile features contribute to identifying different kinds of word usage patterns of individuals, which are not always found in posts. According to S-LIWC, Singaporeans who are high in conservative are likely to use words from the categories such as ``health'', ``home'', ``religion'', and ``family''. Over here, the conservative's positive association with self-transcendence but negative association with self-enhancement may be the reason.

On the whole, our stack model shows the best results almost in all the cases and our models achieve significant improvements, 39.9\% in Conservative, 44.2\% in Hedonism, 54.1\% in Openness to Change, 47.5\% in Self-Transcendence, and 55.4\% in Self-Enhancement, compared to the best of IBM personality and random baselines (when $K$= 50\% for the classification task). 
Our model clearly benefits from using user community specific features. There could be differences in the personal value profile distribution for different user communities. Without considering the distribution differences and word use patterns, the prediction accuracy of personal values can suffer significantly. 

\subsection{Users' Behavior in Twitter vs Personal Values}
In the next study, we apply our personal values prediction model to the analysis of a much larger Singapore social media user population finding the connections between their social media behavior and personal value profiles. Traditionally, such kind of studies could only be carried out by user surveys which were often limited to small number of users due to cost and did not involve real user behavioral data.

We use the Twitter dataset (see Section~\ref{sec:dataset}), consisting of 85,308 users. This is the largest dataset we know that has been used in such a personal values related behavior study. We seek to find out if there are significant relationships between individuals' behavior in Twitter and their personal values determined by our stack model trained for the mid-split prediction task (i.e., $K=50\%$) on the complete Facebook dataset. The stack model uses only post features based on S-LIWC word categories as profile features of Twitter are not the same as those of Facebook. All the tweets of the users are considered to extract post features. To avoid the subsequent behavioral study to be affected by personal values prediction errors, we only consider top and lowest $x$ users ranked by their predicted personal values for each value dimension in the following analysis. $x$ was empirically set to $5000$ to include the 20\% most confident prediction results.  

To check whether the predicted labels are reasonable, Table~\ref{tab:tweet_correlations} shows the correlations among the predicted personal value dimensions of the Twitter users. We clearly observe the correlation of Schwartz' personal values, where consecutive value dimensions are positively correlated (i.e., Conservative vs Self-Transcendence, and Hedonism vs Self-Enhancement) and significant negative correlations among opposite value dimensions (i.e, Self-Enhancement vs Self-Transcendence, and Conservative vs Openness to Change). This observation shows that the stack model predicts values with reasonably good consistency, albeit a lack of ground truth labels for accuracy evaluation.

\begin{table}
\centering
\scriptsize
\caption{Correlations between predicted personal value dimensions for the Twitter dataset (Significant correlations are boldfaced.)}\label{tab:tweet_correlations}
\vspace{-2mm}
\begin{tabular}{|l|c|c|c|c|}
\cline{2-5}
\cline{2-5} 
\multicolumn{1}{c}{} &  
\multicolumn{1}{|c|}{ST} & \multicolumn{1}{c|}{OC} & \multicolumn{1}{c|}{HE} & \multicolumn{1}{c|}{SE} \\ \cline{2-5} \hline
{Conservative (CO)}                                                   & \textbf{0.43}                                 & \textbf{-0.44}                         & \textbf{-0.29}                             & \textbf{-0.29}                                \\ \hline
{Self-Transcendence (ST)}                                                 &                          & \textbf{-0.34}                                   &  \textbf{-0.46}               & \textbf{-0.71}                                \\ \hline
{Openness to Change (OC)}                   &&                                                                              & 0.08                             & 0.08                                 \\ \hline
{Hedonism (HE)}                              &&& &\textbf{0.23}                                      \\ \hline
\end{tabular}
\vspace{-4mm}
\end{table}

\begin{figure*}
\scriptsize
\centering
\hspace{1em}
\subfloat[]{%
\begin{tikzpicture}
\begin{axis}
 [
    ybar=1pt,
    bar width=8pt,
    width=3.9cm,
    height=3cm,
    xtick=data,
    symbolic x coords = {CO,HE,OC,ST,SE},
    enlarge x limits=0.1,
    scaled ticks=false,
    scale only axis,
    legend pos = north west,
    ymin=0
    ]
    
    \addplot+[gray, opacity=0.45, error bars/.cd,
y dir=both,y explicit, error bar style={black}] coordinates{
        (CO,55.81) +- (0.0, 17.49)
        (HE,30.32) +- (0.0, 16.42)
        (OC,13.03) +- (0.0, 5.98)
        (ST,26.40) +- (0.0, 15.64)
        (SE,105.49) +- (0.0, 22.76)};
    \addplot+[black, opacity=0.65, error bars/.cd,
y dir=both,y explicit] coordinates{
        (CO,13.94) +- (0.0, 11.21)
        (HE,10.24) +- (0.0, 5.2)
        (OC,149.31) +- (0.0, 25.12)
        (ST,34.05) +- (0.0, 8.25)
        (SE,48.82) +- (0.0, 12.41)};
    \legend{Low,High}
\end{axis}
\end{tikzpicture}
\label{fig:tot_twt}%
}
\hspace{2em}
\subfloat[]{%
\begin{tikzpicture}
\begin{axis}
    [
    ybar=1pt,
    bar width=8pt,
    width=3.9cm,
    height=3cm,
    xtick=data,
    symbolic x coords = {CO,HE,OC,ST,SE},
    x tick label style={
        /pgf/number format/1000 sep=
    },
    enlarge x limits=0.1,
    scaled ticks=false,
    scale only axis,
    xmode=normal,
    legend pos = north west
    ]
    \addplot+[gray, opacity=0.45,error bars/.cd,
y dir=both,y explicit] coordinates{
        (CO,0.14) +- (0.0, 0.03)
        (HE,0.15) +- (0.0, 0.02)
        (OC,0.14) +- (0.0, 0.08)
        (ST,0.15) +- (0.0, 0.03)
        (SE,0.35) +- (0.0, 0.12)
        };
    \addplot+[black, opacity=0.65,error bars/.cd,
y dir=both,y explicit] coordinates{
        (CO,0.15) +- (0.0, 0.04)
        (HE,0.16) +- (0.0, 0.05)
        (OC,0.30) +- (0.0, 0.05)
        (ST,0.19) +- (0.0, 0.06)
        (SE,0.13) +- (0.0, 0.03)
        };
    \legend{Low,High}
\end{axis}
\end{tikzpicture}
\label{fig:retwt}%
}
\hspace{2em}
\subfloat[]{%
\begin{tikzpicture}
\begin{axis}
    [
    ybar=1pt,
    bar width=8pt,
    width=3.9cm,
    height=3cm,
    xtick=data,
    symbolic x coords = {CO,HE,OC,ST,SE},
    x tick label style={
    },
    enlarge x limits=0.1,
    scaled ticks=false,
    scale only axis,
    xmode=normal,
    legend pos = north west
    ]
    \addplot+[gray, opacity=0.45,error bars/.cd,
y dir=both,y explicit] coordinates{
        (CO,0.48) +- (0.0, 0.06)
        (HE,0.37) +- (0.0, 0.04)
        (OC,0.39) +- (0.0, 0.04)
        (ST,0.46) +- (0.0, 0.04)
        (SE,0.40) +- (0.0, 0.07)}; 
    \addplot+[black, opacity=0.65,error bars/.cd,
y dir=both,y explicit] coordinates{
        (CO,0.33) +- (0.0, 0.03)
        (HE,0.39) +- (0.0, 0.04)
        (OC,0.57) +- (0.0, 0.04)
        (ST,0.37) +- (0.0, 0.03)
        (SE,0.45) +- (0.0, 0.02)};
    \legend{Low,High}
\end{axis}
\end{tikzpicture}
\label{fig:async_rel}%
}
\caption{(a) Average number of tweets by top (high) and bottom (low) ranked $x$ (= 5000) Twitter users in each value dimension; (b) Average \% of retweets by top (high) and bottom (low) ranked $x$ (= 5000) Twitter users in each value dimension; and (c) Average $\frac{\# \;of \;friends}{\# \;of \;followees}$ of top (high) and bottom (low) ranked $x$ (= 5000) Twitter users in each value dimension ((ST=Self-Transcendence, SE=Self-Enhancement, CO=Conservative, OC=Openness to Change, HE=Hedonism), Standard deviations are marked as error bars)}\label{fig:example1}
\vspace{-4mm}
\end{figure*}
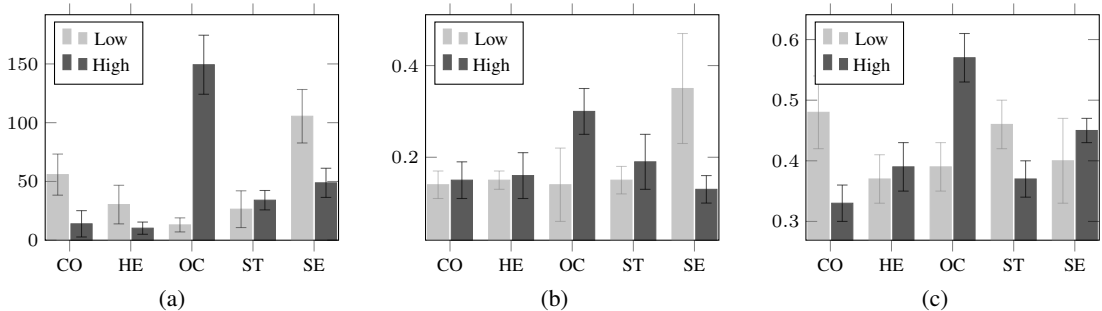

We consider three types of behavior Twitter users demonstrate their activeness.  They are characterized by: (a) number of original tweets generated during $d$ period of time; (b) \% of retweets ($\frac{\#\ of\  retweets}{\#\ of\ tweets}$) during $d$ period of time; and (c) friend-to-follower ratio ($\frac{\#\;of\;friends}{\#\;of\;followers}$). We only report the results of $d$ chosen to be the first month of 2017 as similar results were observed for other $d$ settings. 
This study will seek to confirm a few hypotheses proposed by previous works about the individuals' behaviors in Twitter and their personal values. 

\textbf{Hypothesis 1: High openness individuals are heavy users of Twitter. They tweet more often than other individuals \cite{sumner2012predicting}.} Our results in Figure~\ref{fig:tot_twt} clearly supports the hypothesis. Users high in openness to change generate more tweets on average than the users higher in other value dimensions. Figure~\ref{fig:tot_twt} also shows that users low in openness to change are less active, with a small average number of original tweets. Moreover, the individuals high in Hedonism have the least average number of original tweets, which also indirectly supports a claim made in \cite{marshall2018intellectual} stating that extroverts (usually higher in hedonism) prefer to use other social media sites like Facebook for social purposes while Twitter is often used for informational purposes.

\textbf{Hypothesis 2: Control and dominance over others as well as expression of personal interests are common among individuals who are high in Self-Enhancement. \cite{schwartz2012overview}.} As shown in Figure~\ref{fig:retwt}, Twitter users high in self-enhancement show very small \% of retweets, suggesting that they do not like to share other users' opinions. In contrast, users with very low self-enhancement are very active in retweeting. This observation clearly supports Hypothesis 2 which states that individuals high in self-enhancement want to control and dominate others instead of following others' opinions.

\textbf{Hypothesis 3: Protection of order and harmony in their relationships, and selective in making relationships are common among conservative people \cite{schwartz1992universals}.}
Figure~\ref{fig:async_rel} shows that there is a significant asymmetry in their friend-to-followee ratio. On average, only 33\% of followers are getting followed for Twitter users with high conservative values.  In contrast, users with low conservative values enjoy a higher friend-to-follower ratio (around 0.48).  This observation indicates that conservative users prefer to select their social network carefully. Moreover, Figure~\ref{fig:async_rel} shows that high openness users like to make more mutual friends in Twitter compared to the users high in other value dimensions. This further supports to the fact that their tendency to follow new information and being active in Twitter, which is stated in Hypothesis 1.

\textbf{Summary of Findings.}  Despite the differences between Twitter and Facebook, this study shows that the prediction models trained using the latter can be used to derive interesting findings of behavior demonstrated by Twitter users with different personal values. Note that this study has carefully selected users from the same community (i.e., social media users in Singapore), similar findings may not be replicated when the prediction models are trained using data from a completely unrelated user population. One possible approach to address this mismatch of training and test data is to adopt transfer learning to adapt the prediction models, which is another topic of research that should be studied in the future \cite{calais2011bias}.

\subsection{Discussion}
\textbf{Ethical Concerns. }To protect the privacy of the participants when collecting and handling the data in this task, we followed a research protocol approved by the Institutional Review Board (IRB) of the authors’ university. Also, the datasets were anonymized before using them for our model.

To avoid the privacy concerns related to the applications, our model is designed such that the personal values of a given user is predicted merely based on the user's text content (i.e., without using any shared features). Our model therefore could be deployed at the users' end, instead of in a central system. The user's postings can be used to predict the personal values without having to be shipped to a central system. Such personal value descriptors stored at the user end could already enable many useful services. For example, they could be used to personalize recommendations (e.g., suitable jobs) proposed to users via different recommendation engines based on the users' personal values~\cite{silva2020jplink,kern2019social}. 

\textbf{Generalizability. }In this work, the language feature generation using community-specific LIWC (i.e., S-LIWC) is the only part specific to Singapore users. However, we observe that the proposed model outperforms the baselines with even conventional LIWC too. Thus, the proposed model is generalizable for other datasets. Also, the proposed model could be integrated with any community-specific LIWC to generate community-specific predictions. 

\section{Conclusion}
\label{sec:conclude}
In this paper, we first study how personal values prediction using a user's social media content can be significantly improved by considering geographic differences in word usage and profile information. 

In addition, we proposed a new stack model to predict individuals' personal values by exploiting the correlations between personal values. 
Through our experiments, we show a significant boost in prediction accuracy for our proposed stack model compared to the state-of-the-art models proposed in previous works (e.g., IBM Personality Insight API). 
We finally showed that our model predicts personal values of a large set of Twitter users and derived interesting findings linking their personal values to their behavior on Twitter. These findings are largely consistent with previous research using traditional survey based studies. 
With reasonably accurate personal values prediction models, we envisage that many interesting research studies on the impact of personal values to opinion polarization, community formation, and others can be carried out at scale. 
Although we have considered the individual's word usage in user profiles, i.e., groups, likes, and interests, there are other behavioral features that can be used to enhance the prediction accuracy of personal values which should be studied in future work. In addition, incorporating network structure and dynamic nature in social media to predict personal values may also be another promising future direction. 

\bibliographystyle{aaai}
\bibliography{ICWSM_references_short.bib}

\begin{thebibliography}{}

\bibitem[\protect\citeauthoryear{Boyd \bgroup et al\mbox.\egroup
  }{2015}]{boyd2015values}
Boyd, R.~L.; Wilson, S.~R.; Pennebaker, J.~W.; Kosinski, M.; Stillwell, D.~J.;
  and Mihalcea, R.
\newblock 2015.
\newblock Values in words: Using language to evaluate and understand personal
  values.
\newblock In {\em ICWSM}.

\bibitem[\protect\citeauthoryear{Braithwaite and
  Law}{1985}]{braithwaite1985structure}
Braithwaite, V.~A., and Law, H.
\newblock 1985.
\newblock Structure of human values: Testing the adequacy of the rokeach value
  survey.
\newblock {\em Journal of personality and social psychology}.

\bibitem[\protect\citeauthoryear{Calais~Guerra \bgroup et al\mbox.\egroup
  }{2011}]{calais2011bias}
Calais~Guerra, P.~H.; Veloso, A.; Meira~Jr, W.; and Almeida, V.
\newblock 2011.
\newblock From bias to opinion: a transfer-learning approach to real-time
  sentiment analysis.
\newblock In {\em KDD}.

\bibitem[\protect\citeauthoryear{Caprara \bgroup et al\mbox.\egroup
  }{2006}]{caprara2006personality}
Caprara, G.~V.; Schwartz, S.; Capanna, C.; Vecchione, M.; and Barbaranelli, C.
\newblock 2006.
\newblock Personality and politics: Values, traits, and political choice.
\newblock {\em Political psychology}.

\bibitem[\protect\citeauthoryear{Chen \bgroup et al\mbox.\egroup
  }{2014}]{chen2014understanding}
Chen, J.; Hsieh, G.; Mahmud, J.~U.; and Nichols, J.
\newblock 2014.
\newblock Understanding individuals' personal values from social media word
  use.
\newblock In {\em CSCW}.

\bibitem[\protect\citeauthoryear{Cronbach}{1951}]{cronbach1951coefficient}
Cronbach, L.~J.
\newblock 1951.
\newblock Coefficient alpha and the internal structure of tests.
\newblock {\em psychometrika}.

\bibitem[\protect\citeauthoryear{Golbeck, Robles, and
  Turner}{2011}]{golbeck2011predicting}
Golbeck, J.; Robles, C.; and Turner, K.
\newblock 2011.
\newblock Predicting personality with social media.
\newblock In {\em CHI}.

\bibitem[\protect\citeauthoryear{Grankvist and
  Kajonius}{2015}]{grankvist2015personality}
Grankvist, G., and Kajonius, P.
\newblock 2015.
\newblock Personality traits and values: a replication with a swedish sample.
\newblock {\em International Journal of Personality Psychology}.

\bibitem[\protect\citeauthoryear{Hofstede}{1984}]{hofstede1984culture}
Hofstede, G.
\newblock 1984.
\newblock {\em Culture's consequences: International differences in
  work-related values}.
\newblock sage.

\bibitem[\protect\citeauthoryear{Hsieh \bgroup et al\mbox.\egroup
  }{2014}]{hsieh2014you}
Hsieh, G.; Chen, J.; Mahmud, J.~U.; and Nichols, J.
\newblock 2014.
\newblock You read what you value: understanding personal values and reading
  interests.
\newblock In {\em CHI}.

\bibitem[\protect\citeauthoryear{Huberman, Romero, and
  Wu}{2009}]{huberman2008social}
Huberman, B.~A.; Romero, D.~M.; and Wu, F.
\newblock 2009.
\newblock {social networks that matter: Twitter under the microscope}.
\newblock {\em First Monday}.

\bibitem[\protect\citeauthoryear{Inglehart}{1997}]{inglehart1997modernization}
Inglehart, R.
\newblock 1997.
\newblock {\em Modernization and postmodernization: Cultural, economic, and
  political change in 43 societies}.
\newblock Princeton University Press.

\bibitem[\protect\citeauthoryear{Jin}{2013}]{jin2013peeling}
Jin, S.-A.~A.
\newblock 2013.
\newblock {Peeling back the multiple layers of Twitter’s private disclosure
  onion: The roles of virtual identity discrepancy and personality traits in
  communication privacy management on Twitter}.
\newblock {\em New Media \& Society}.

\bibitem[\protect\citeauthoryear{Kern \bgroup et al\mbox.\egroup
  }{2019}]{kern2019social}
Kern, M.~L.; McCarthy, P.~X.; Chakrabarty, D.; and Rizoiu, M.-A.
\newblock 2019.
\newblock {Social media-predicted personality traits and values can help match
  people to their ideal jobs}.
\newblock {\em PNAS}.

\bibitem[\protect\citeauthoryear{Maheshwari \bgroup et al\mbox.\egroup
  }{2017}]{maheshwari2017societal}
Maheshwari, T.; Reganti, A.~N.; Gupta, S.; Jamatia, A.; Kumar, U.; Gamb{\"a}ck,
  B.; and Das, A.
\newblock 2017.
\newblock A societal sentiment analysis: Predicting the values and ethics of
  individuals by analysing social media content.
\newblock In {\em EACL}.

\bibitem[\protect\citeauthoryear{Maio}{2010}]{maio2010mental}
Maio, G.~R.
\newblock 2010.
\newblock Mental representations of social values.
\newblock In {\em Advances in experimental social psychology}. Elsevier.

\bibitem[\protect\citeauthoryear{Marshall \bgroup et al\mbox.\egroup
  }{2018}]{marshall2018intellectual}
Marshall, T.~C.; Ferenczi, N.; Lefringhausen, K.; Hill, S.; and Deng, J.
\newblock 2018.
\newblock {Intellectual, narcissistic, or Machiavellian? How Twitter users
  differ from Facebook-only users, why they use Twitter, and what they tweet
  about}.
\newblock {\em Psychology of Popular Media Culture}.

\bibitem[\protect\citeauthoryear{Mikolov \bgroup et al\mbox.\egroup
  }{2013}]{mikolov2013distributed}
Mikolov, T.; Sutskever, I.; Chen, K.; Corrado, G.~S.; and Dean, J.
\newblock 2013.
\newblock Distributed representations of words and phrases and their
  compositionality.
\newblock In {\em NIPS}.

\bibitem[\protect\citeauthoryear{Misra \bgroup et al\mbox.\egroup
  }{2016}]{misra2016cross}
Misra, I.; Shrivastava, A.; Gupta, A.; and Hebert, M.
\newblock 2016.
\newblock {Cross-stitch networks for multi-task learning}.
\newblock In {\em CVPR}.

\bibitem[\protect\citeauthoryear{Mukta, Ali, and
  Mahmud}{2017}]{mukta2017identifying}
Mukta, M. S.~H.; Ali, M.~E.; and Mahmud, J.
\newblock 2017.
\newblock Identifying and predicting temporal change of basic human values from
  social network usage.
\newblock In {\em ASONAM}.

\bibitem[\protect\citeauthoryear{Mukta \bgroup et al\mbox.\egroup
  }{2017}]{md2017predicting}
Mukta, M. S.~H.; Khan, E.~M.; Ali, M.~E.; and Mahmud, J.
\newblock 2017.
\newblock Predicting movie genre preferences from personality and values of
  social media users.
\newblock In {\em ICWSM}.

\bibitem[\protect\citeauthoryear{Parks-Leduc, Feldman, and
  Bardi}{2015}]{parks2015personality}
Parks-Leduc, L.; Feldman, G.; and Bardi, A.
\newblock 2015.
\newblock Personality traits and personal values: A meta-analysis.
\newblock {\em Personality and Social Psychology Review}.

\bibitem[\protect\citeauthoryear{Pennebaker \bgroup et al\mbox.\egroup
  }{2015}]{pennebaker2015development}
Pennebaker, J.~W.; Boyd, R.~L.; Jordan, K.; and Blackburn, K.
\newblock 2015.
\newblock The development and psychometric properties of liwc2015.
\newblock Technical report.

\bibitem[\protect\citeauthoryear{Rokeach}{1973}]{rokeach1973nature}
Rokeach, M.
\newblock 1973.
\newblock {\em The nature of human values.}
\newblock Free press.

\bibitem[\protect\citeauthoryear{Schwartz}{1992}]{schwartz1992universals}
Schwartz, S.~H.
\newblock 1992.
\newblock Universals in the content and structure of values: Theoretical
  advances and empirical tests in 20 countries.
\newblock In {\em Advances in experimental social psychology}. Elsevier.

\bibitem[\protect\citeauthoryear{Schwartz}{2003}]{schwartz2003proposal}
Schwartz, S.~H.
\newblock 2003.
\newblock A proposal for measuring value orientations across nations.
\newblock {\em Questionnaire Package of the European Social Survey}.

\bibitem[\protect\citeauthoryear{Schwartz}{2012}]{schwartz2012overview}
Schwartz, S.~H.
\newblock 2012.
\newblock An overview of the schwartz theory of basic values.
\newblock {\em Online readings in Psychology and Culture}.

\bibitem[\protect\citeauthoryear{Silva, Lo, and Lim}{2020}]{silva2020jplink}
Silva, A.; Lo, P.-C.; and Lim, E.-P.
\newblock 2020.
\newblock {JPLink: On Linking Jobs to Vocational Interest Types}.
\newblock In {\em PAKDD}.

\bibitem[\protect\citeauthoryear{Sumner \bgroup et al\mbox.\egroup
  }{2012}]{sumner2012predicting}
Sumner, C.; Byers, A.; Boochever, R.; and Park, G.~J.
\newblock 2012.
\newblock {Predicting dark triad personality traits from Twitter usage and a
  linguistic analysis of tweets}.
\newblock In {\em ICMLA}.

\bibitem[\protect\citeauthoryear{Verplanken and
  Holland}{2002}]{verplanken2002motivated}
Verplanken, B., and Holland, R.~W.
\newblock 2002.
\newblock Motivated decision making: Effects of activation and self-centrality
  of values on choices and behavior.
\newblock {\em Journal of personality and social psychology}.

\end{thebibliography}

\end{document}